\documentclass[review, 1p, 11pt]{elsarticle}

\usepackage{graphicx}
\usepackage{psfrag}
\usepackage{subfigure}
\usepackage{stfloats}
\usepackage{amsmath}
\usepackage{amsmath}
\usepackage{amssymb}
\usepackage{amsthm}
\usepackage{color}

\def\1{\,\mathds{1}\,}

\def\I{{\cal I}}

\def\x{{\bf x}}

\def\I-mode{\textit{I-mode}}
\def\S-mode{\textit{S-mode}}

\numberwithin{equation}{section}

\graphicspath{{figures/}}

\begin{document}

\begin{frontmatter}

\title{MAC-aware Routing Metrics for the Internet of Things}
\fntext[t1]{P. Mekikis was with the School of Electrical Engineering, KTH, when contributing to this work.}
\author[kth]{P. Di Marco\corref{cor1}}
\ead{pidm@kth.se}
\author[kth]{G. Athanasiou}
\ead{georgioa@kth.se}
\author[upc]{P.-V. Mekikis\fnref{t1}}
\ead{vmekikis@tsc.upc.edu}
\author[kth]{C. Fischione}
\ead{carlofi@kth.se}
\address[kth]{ACCESS Linnaeus Centre, School of Electrical Engineering, KTH, Stockholm, Sweden}
\address[upc]{Signal Theory and Comm. Dept., Technical University of Catalonia, Barcelona, Spain}
\cortext[cor1]{Corresponding author}


\begin{abstract}

Wireless medium access control (MAC) and routing protocols are
fundamental building blocks of the Internet of
Things (IoT). As new IoT networking standards are being proposed and different
existing solutions patched, evaluating the end-to-end performance of the
network becomes challenging. Specific solutions designed to be
beneficial, when stacked may have detrimental effects on the
overall network performance. In this paper, an analysis of MAC and routing protocols for IoT is provided with focus
 on the IEEE 802.15.4 MAC and the IETF RPL standards. It is shown that existing routing metrics do not
account for the complex interactions between MAC and routing, and thus novel metrics are proposed.
This enables a protocol selection mechanism  for selecting the routing option and adapting the MAC parameters,
given specific performance constraints. Extensive analytical and experimental results show that the behavior of the
MAC protocol can hurt the performance of the routing protocol and vice versa, unless these two are carefully
optimized together by the proposed method.
\end{abstract}
\end{frontmatter}

\section{Introduction}

 Internet of Things (IoT) has been introduced in 1999 to
envision the concept of connecting physical
objects, such as sensors and smart phones, supporting direct wireless connection to
the internet. This enables an immense variety of applications in an Internet-like framework~\cite{Atzori}.
Any terminal connected to the Internet will be able to interact with these
objects. Applications include among others, building and industrial automation,
health-care, personal wireless communications, smart grids, and
security. However, to support such an increasing
number of emerging applications, the wireless medium access
control (MAC) and the routing protocols must be inherently scalable,
interoperable, and they must have a solid standardization base to support
future innovations.

Differently from classic wireless ad hoc and sensor network applications,
IoT applications have complex and heterogeneous requirements.
The network performance is not only measured in terms of throughput, but
other indicators, such as reliability, delay, and energy consumption, must be
jointly optimized and adapted to different application scenarios~\cite{trend}.
Assuming a network of several objects interconnected to support an IoT application
where efficient information routing is essential, the decision over different routing paths
highly depends on link performance indicators, which are  influenced by the MAC parameters.
On the other side, the routing determines the distribution of the traffic load in the
network, which significantly affects the aforementioned performance indicators.
Therefore, it is inefficient to design MAC and routing protocols separately and
their joint adaptation is essential to achieve the desired network performance.
The essential question is which are the metrics that enable such an efficient joint adaptation.

By following a classic layered design process, standardization bodies such as IEEE and Internet Engineering Task Force (IETF) are
working independently on the design of the future MAC and routing protocols for IoT.
According to recent surveys~\cite{palattella_iot}, the protocol stack for
IoT applications integrates the IEEE 802.15.4~\cite{ieee802154} standard for physical and MAC layers,
and the IPv6 routing protocol for low power and lossy networks
(RPL)~\cite{ROLL}.
The IEEE 802.15.4 defines flexible physical and MAC layers
for low data rate and low power applications.
The standard has been adopted with
some modification also by a number of other solutions, including
ZigBee, WirelessHART, ISA-100~\cite{BarontiCC}, and it
already represents more than 50\% of the building and industrial
automation market~\cite{zigbee2011}.
At the networking layers, the
IPv6 routing protocol for low power and lossy networks
(RPL) represents the reference standard proposed by the IETF for IPv6-compatible IoT applications.

In this paper, we first emphasize the main characteristics of the IEEE 802.15.4 MAC and the IETF RPL protocols. We then present an analysis to
characterize the protocols mutual effects and their dynamics. In particular, we show that the level of contention at the MAC layer
impacts the routing decisions in an unexpected manner. In addition, we show that in the presence of dominant paths
in the network, i.e., paths with high traffic forwarding, the performance indicators at the MAC layer are significantly affected. Based on our analysis, we propose metrics that guide the interactions between MAC and routing, and we introduce a mechanism that selects the appropriate routing metric and adapts the corresponding MAC parameters to fulfill one of the most important requirement of several IoT applications: minimize the energy consumption in the network, given reliability and delay constraints.

Our methodology starts from the modeling of the behavior of the current protocols, motivating in this ways the design of new efficient metrics. The proposed metrics are numerically and experimentally validated in a realistic environment and in comparison to existing metrics. To the best of our knowledge this is the first paper to present in depth modeling of the protocol interactions, to propose cross layer routing metrics supported by a protocol selection and adaptation mechanism, and to give an extensive experimental study where approaches in literature are compared. We believe that the outcome of this work could give important insights to influence the standardization process for the IoT.

The rest of the paper is organized as follows. In Section~\ref{sec:rel}, we survey the related literature. Section~\ref{sec:over-uns}, gives an overview of the basic functionalities/parameters of the IEEE 802.15.4 MAC and the IETF RPL protocols. In Section~\ref{sec:complex-relation}, we motivate our analysis and emphasize the importance of cross-layer interactions, considering the requirements of IoT applications. In
Section~\ref{sec:interactions}, the interactions among MAC and
routing are modeled. Then, Section~\ref{sec:mac-aware} presents novel MAC-aware routing metrics, following the lines of our analysis.  In Section~\ref{exp:res}, we present the experimental evaluation of the proposed metrics in comparison to existing approaches. Finally, Section~\ref{sec:conclusions}
concludes the paper.

\section{Related Work}\label{sec:rel}

The importance of a mathematical modeling of MAC protocols related to sensor networking has been advocated in recent literature~\cite{misicCC}\nocite{Park_TON}~--~\cite{Sinem09_glo}. The critical effect of MAC parameters on the performance has been shown and discussed in~\cite{misicCC}  for IEEE 802.15.4 networks.
In~\cite{Park_TON}, a Markov chain model is used to design a distributed adaptive algorithm for minimizing the power consumption of single hop star networks using the IEEE 802.15.4 MAC, while guaranteeing a given successful packet reception
probability and delay constraints.
In~\cite{Sinem09_glo}, an automatic MAC protocol selection mechanism is proposed.
The main idea of this approach is to provide a mathematical analysis of various MAC protocols, including the IEEE 802.15.4 MAC,
and to choose the optimal MAC protocol and adapt its parameters for the selected modality, topology, and packet
generation rate. In particular the designed mechanism takes into account the corresponding physical
layer technology and hardware, while satisfying constraints for energy, reliability, and delay.
The value of the aforementioned approaches is that the algorithms do not require any modification of the IEEE 802.15.4
standard to be applied.
However, their application is limited to single-hop networks.

Continuing our literature review in the MAC design and adaptation, a framework for MAC parameter adaptation based on analytical modeling has been also presented
in~\cite{ptunes}. This approach has been developed for XMAC and LPP protocol, but the results are not intended for the IEEE 802.15.4 MAC. An adaptation mechanism of the IEEE 802.15.4 MAC has been proposed in~\cite{Timmons04}. The mechanism was experimentally validated in a wireless body sensor network deployed for medical applications. However, this approach does not include any analytical modeling of the performance of the MAC and the adaptations are performed based on observations during their experimental study.

Within the literature related to the study of the routing protocols for IoT applications, \cite{perf_RPL} presents an experimental performance evaluation
of RPL using the basic hop count routing metric and the expected transmissions count (ETX)~\cite{ETX} metric.
The ETX is a reliability metric that indicates the number of retransmissions a node expects to execute to successfully deliver a packet  to the
destination node.
However, the study in \cite{perf_RPL} does not consider the performance of RPL when a contention-based MAC protocol is active and there is no proposal about new routing metrics.
In the back-pressure collection protocol (BCP)~\cite{BCP}, an extension of the ETX metric lead to the introduction of a dynamic back-pressure routing metric. In BCP, the routing and forwarding decisions are made on a per-packet basis by computing a back-pressure weight of each outgoing
link that is a function of the queues of the nodes and the link state information. BCP is tested over a low power contention-based MAC. However, the effects of the limited number of backoffs and retransmissions, present in the IEEE 802.15.4 standard, are not taken into account.

In~\cite{ghadimi_secon}, the authors propose a metric for opportunistic routing for very low-duty cycled MACs that considers the expected number required to successfully deliver a packet from source to destination. In~\cite{Pavkovic:2011}, a multi-path opportunistic routing is proposed for time-constrained operations over IEEE 802.15.4 MAC. However, load balancing and the effects of contention-based access are not considered in both previous approaches.
A study on the interaction of RPL with the MAC layer is presented in~\cite{RPL11}, where  the use of a receiver-initiated MAC protocol in enhancing the performance of RPL is investigated.
In~\cite{Han13}, a cross-layer framework has been proposed for IoT applications, by considering SMAC and RPL.
However, these works do not take into account the specifications of the IEEE 802.15.4 MAC, which is widely used as the default MAC protocol for IoT~\cite{palattella_iot}.

Our work differentiates from the aforementioned approaches by modeling the IEEE 802.15.4 MAC and RPL protocol interactions, when applied in multi-hop networks.  Considering complex multi-hop topologies, in comparison to single hop topologies of previous studies, we investigate the performance of the realistic network deployments for supporting IoT applications. In particular, we propose cross layer routing metrics supported by a protocol selection and adaptation mechanism. Finally, we present an extensive experimental validation of our approach in comparison to other routing metrics.

\section{Protocols Overview}\label{sec:over-uns}

In this section, we give an overview of the main functionalities of the IEEE 802.15.4 MAC and RPL protocols, which allows us to characterize their performance later on in the next sections.

\subsection{IEEE 802.15.4 MAC Protocol}\label{sec:mac}

The  IEEE 802.15.4 MAC defines two basic access
modalities: contention-based MAC, with a simple unslotted
carrier sense multiple access collision avoidance (CSMA/CA), and
hybrid-based MAC with a slotted CSMA/CA
and a contention-free operation based on guaranteed time slot (GTS) allocation.
In the following, we focus on the unslotted modality, which is of major interest for RPL.

Consider a node $V_i$ trying to transmit a packet.
First, the node waits for a random number of time units in the window $[0 - 2^{m_0}]$. Then, the
node performs a clear channel assessment (CCA). If the channel is
idle, the node begins the packet transmission. In case that the CCA
fails due to busy channel, the MAC layer increases the backoff window exponentially.
We indicate by $\alpha_i$ the probability of busy channel for node $V_i$. If the backoff exponent
reaches a maximum value $m_b$, it remains at that value until it is reset. If the number of backoffs exceeds
a maximum number $m$, then the packet is discarded due to channel access failures.
Otherwise the CSMA/CA algorithm
generates a random number of complete backoff periods and repeats
the process.
The reception of an acknowledgement (ACK) is interpreted as successful packet
transmission. If the node fails to receive the ACK, the MAC re-initializes
the backoff window and follows the CSMA/CA mechanism to re-access
the channel. After a maximum number of retransmissions $n$, the packet is discarded.

\subsection{Routing Protocol for Low Power and Lossy Networks (RPL)}\label{sec:routing-protocols}
RPL constructs destination-oriented directed acyclic graphs
(DODAGs) over the network, according to optimization objectives.
Every node in a DODAG is identified by a rank, a scalar value
that represents the relative position of the node with respect to
other nodes and the DODAG root.
Nodes build and maintain DODAGs by periodically multicasting
messages, called DODAG information object (DIO), to their
neighbors. To join a
DODAG, a node listens to the DIO messages sent by its neighbors,
selects a subset of these nodes as its parents, and compute its rank.
Although the rank is computed by using link costs, topology building and
maintenance mechanisms can be made independently of packet
forwarding procedures.

\begin{figure}\centering
  \includegraphics[width=0.9\textwidth]{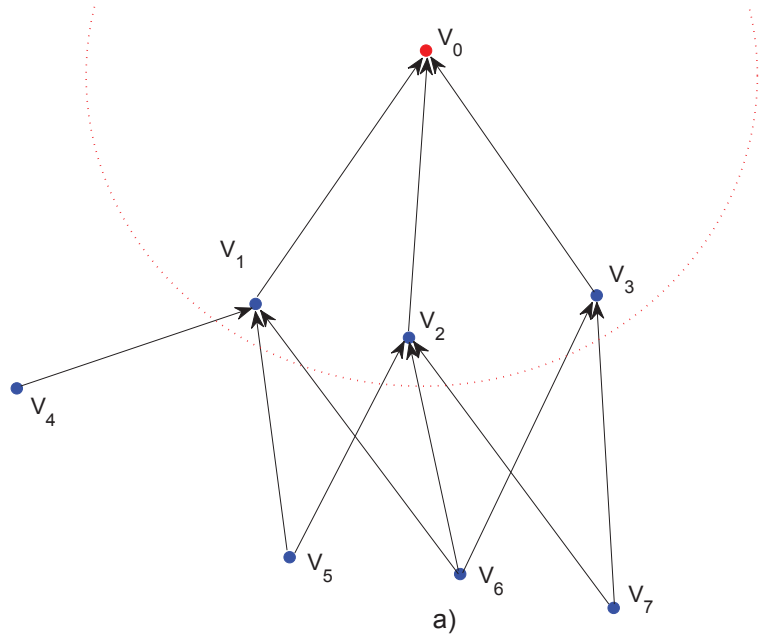}
  \includegraphics[width=0.9\textwidth]{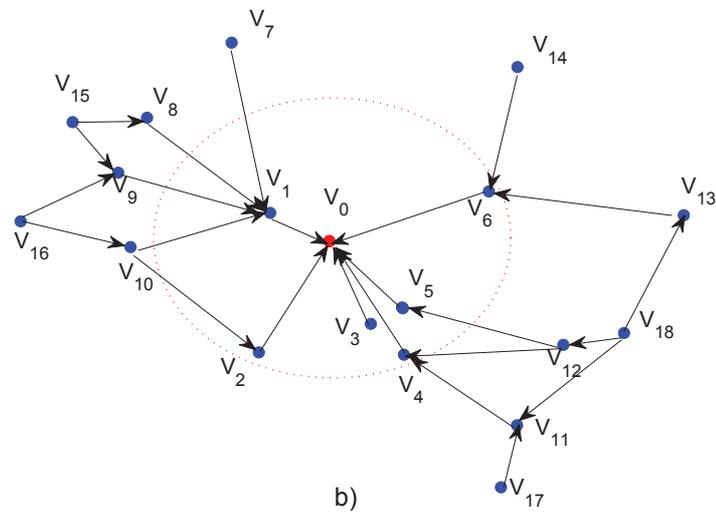}
  \caption{DODAG example for a randomly generated topology with 7 nodes (a) and 18 nodes (b).}
  \label{fig:topology}
\end{figure}

In Fig.~\ref{fig:topology}, we visualize two randomly generated DODAG examples that we use as reference topologies in our evaluation study.
Each end-device has one or more nodes in its parent set.
The packet forwarding is based on end-to-end metrics and application
constraints. There are various metrics and constraints that can be
used. Reliability, packet delay, and node energy consumption are
indicators that can be used both as metrics and constraints.
For example in Fig.~\ref{fig:topology}a, we consider node $V_5$ that has $V_1$, $V_2$, and $V_3$ in its parent set.
The metric of a generic link $(i,j)$ is denoted as  $\pi_{i,j}$. Then, $V_5$ chooses $V_1$ as next hop
node if $\pi_{5,1}+\pi_{1,0}$ is better than $\pi_{5,2}+\pi_{2,0}$ and $\pi_{5,3}+\pi_{3,0}$ .
We provide details of the MAC-routing interactions in the next section.

\section{MAC and Routing Interactions}~\label{sec:complex-relation}
\begin{figure}\centering
  \includegraphics[width=0.9\linewidth]{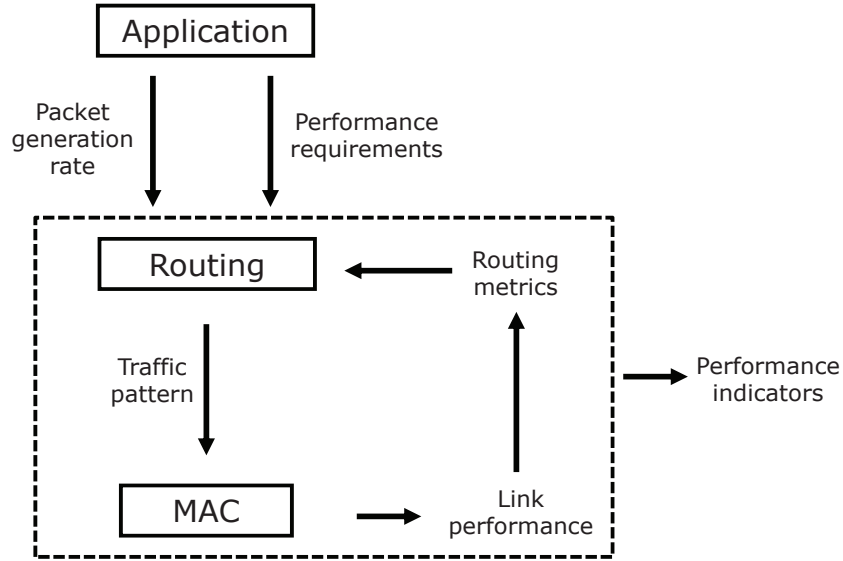}
  \caption{The loop of MAC and routing interactions.
  The closed loop can lead to instabilities if these interactions are not considered when selecting the protocols.}\label{fig:MAC_routing_app_c3}
\end{figure}

In this section, we describe the interactions between MAC and routing through the
feedback loop visualized in Fig.~\ref{fig:MAC_routing_app_c3} and we highlight the importance of cross-layer interactions
in IoT standards.

The supported \emph{application} sets a traffic generation rate
for each node in the network, which is related to the required
sampling time of the sensing operation in case of sensor networks, or a
generic data generation rate of the application. Moreover, the application
layer determines performance requirements for the lower protocol
layers (e.g., minimum data delivery rate and maximum packet delay). The \emph{routing} layer combines
the topological information in a
network communication graph. In addition, based on specific metrics, the routing
protocol takes appropriate decisions to distribute the traffic in the network. It affects the load that
each link, regulated by the \emph{MAC} layer, has to serve. As an output of the
MAC process, we get the link performance in terms of energy consumption,
reliability, or delay. Furthermore, the link performance indicators
may influence directly the routing metric, so closing the loop
between MAC and routing layers. The combined analysis of the MAC and the routing
layers determines the end-to-end performance indicators.

Due to the inter-dependency, if the routing is designed without taking into account the
MAC, the end-to-end
performance indicators could be far from the desired by the
application.
Similar performance anomalies happen when MAC protocol standards are designed or
selected regardless of the applied routing protocol. The wireless channel is
shared, thus a choice of a MAC protocol may
cause unexpected packet collisions or interference over
other links.
We advocate a protocol design methodology for which the cross-layer
interactions are mathematically described as a function of the
MAC and routing parameters, namely the free parameters
that can be regulated to influence the protocol performance. The mathematical description must be accurate enough to be of practical use in adaptation mechanisms for the real world protocols.
By extending the approach introduced in~\cite{Sinem09_glo}, we propose
a mathematical selection tool based on a library of compatible
MAC and routing protocols. The tool picks the best
protocol solution in the library, with optimal parameters for the
specific application. The selection takes as input
the network topology and packet generation rate, the application
requirements for reliability, delay, and energy consumption, and
the resource constraints from the underlying physical layer. Then, the
tool chooses the protocol with optimal parameters for the specific topology and packet
generation rate that gives the best performance. The idea of analytical
description and optimization of the cross layer interactions is
not new, but as a matter of fact, the standardization process seems to consider it only to some little extent.

In the next section, we give a detailed description of the proposed joint MAC and routing model.

\section{Joint MAC and Routing Model} \label{sec:interactions}
This section describes the joint model of the IEEE 802.15.4 MAC and the IETF RPL.

As introduced in Section~\ref{sec:mac}, the level of contention at node $V_i$ is described by the busy channel probability $\alpha_i$, which is a function of the MAC parameters ($m_0$, $m_b$, $m$, and $n$), of the traffic $Q_i$, and of the busy channel probabilities of the neighboring nodes $\alpha_k$.
A model based on Markov chain analysis to derive $\alpha_i$ analytically is presented in~\cite{PG_TVT}. To provide a simple solution, we assume that $\alpha_i$ is estimated at node $V_i$ during the CCA. The busy channel probability is initialized at the beginning of the node's operation. The estimation of the probability uses a sliding window. When the node senses the channel, the probability is updated by $\alpha_i(t) =
r \alpha_i(t-1) + (1-r)\hat{\alpha}_i(t)$ for some $r\in (0,1)$, respectively. Note that $\hat{\alpha}_i(t)$ is the busy channel probability of the current sliding window.

Given that the channel is idle during CCA, the probability $\gamma_{i,j}$ that a packet is not received at destination  is given by
\begin{align} \label{eq:gamma}
\gamma_{i,j} = p_{\mathrm{coll},i} + (1- p_{\mathrm{coll},i}) p_{i,j} \,,
\end{align}
where $p_{\mathrm{coll},i}$ is the probability that $V_i$ encounters a collision due to a simultaneous transmission as the probability that at least another node performs the CCA and finds the channel idle in the same time unit, namely
$p_{\mathrm{coll},i} = \alpha_i/T_s$.
The bad channel probability $p_{i,j}$ is the probability that the link quality is not sufficient to yield a successful packet reception at the receiver $V_j$.
We notice that the probability $\gamma_{i,j}$ is an upper bound of the packet loss probability since it considers unsuccessful all the events in which at least another node is transmitting, independently of the received SINR.

We evaluate the reliability as
the successful packet reception rate or, equivalently, the
complement of the discard probability.
Packets are
discarded due to the following reasons: (i) channel access failure or (ii) retransmissions limit. Channel access failure
happens when a packet fails to obtain clear channel within the maximum number of backoffs $m+1$.
In the assumption of independent channel conditions for consecutive channel accesses, the probability
of finding the channel busy for $m+1$ attempts is $\alpha_i^{m+1}$. This probability has to be considered also for any of the  $n$
retransmission stages. The probability of being in the $k$-th retransmission stage is $ (\gamma_{i,j}(1-\alpha_i^{m+1}))^k$.
Therefore, the probability that the packet is discarded due to channel access failure is
\begin{align} \label{eq:channel_fail}
p^{cf}_{i,j} & = \alpha_i^{m+1} \sum_{k=0}^{n} (\gamma_{i,j}(1-\alpha_i^{m+1}))^k \,
\end{align}

Furthermore, a packet is discarded if the transmission
fails due to repeated collisions and losses after the maximum number of retransmissions $n$.
The probability of a packet being discarded due to the retransmissions limit is
\begin{align} \label{eq:collision_repeat}
p^{cr}_{i,j}  =(\gamma_{i,j}(1-\alpha_i^{m+1}))^{n+1}\,.
\end{align}

The reliability of the link $(i,j)$ is given by
\begin{align} \label{eq:reliability4}
R_{i,j} = 1 - p^{cf}_{i,j} - p^{cr}_{i,j}\,.
\end{align}
We notice that the reliability is dependent only on the busy channel probability $\alpha_i$, the bad channel probability $p_{i,j}$, and the MAC parameters $m,n$.
As explained before, the busy channel probability $\alpha_i$ can be estimated at the transmitter during the CCA.
Therefore, a node does not require any extra communication and sensing state to estimate this probability compared to the default IEEE 802.15.4 standard.
As shown in~\cite{gomez}, the bad channel probability $p_{i,j}$ can be retrieved by the link quality indicator (LQI) which is a parameter offered by the IEEE 802.15.4 physical layer header for every received packets~\cite{ieee802154}.

We consider the delay for
successfully received packets as sum of the queueing time and
the service time. As the service time we consider the time interval from the
instant the packet is ready to be transmitted, until the ACK is received. If a packet is dropped,
its delay is not included into the derivation.
Eventually, we consider the power
consumption of nodes in the network, as the sum of the contributions in backoff, carrier sense,
transmission, reception and idle-queue state. We assume that relay
nodes are in idle-listening state also during the inactivity
period (depending on the duty cycle policy).
We provide more details on the analytic derivation of these performance requirements in~\cite{PG_TVT}.

We represent the dynamical interaction of MAC and routing
using a statistical model.  Let $\lambda_j$ be the
traffic generation rate of node $V_j$. In addition to $\lambda_j$, node $V_j$ has to forward
traffic generated by its children.
Let $Q_j$ be the traffic generated plus the traffic that the node has to
forward (generated and routed by other nodes) to a node in the set of candidate receivers $\Gamma_i$.

%
%

According to the RPL specifications, RPL path selection is modeled by a real valued matrix
$\textbf{M}$, in which element $M_{i,j}$ corresponds to the
probability that the metric $\pi_{i,j}$ evaluated for the link $(i,j)$ is the
best among the set of candidate receivers $\Gamma_i$.


The distribution of the traffic flows along the network is modeled
by the matrix $\textbf{M}$, and by a scaling due to that only
successfully received packets are forwarded. Therefore, we define
a matrix $\textbf{T}$ such that $T_{i,j}= M_{i,j} R_{i,j}$ where
$R_{i,j}$ is the reliability in the link $(i,j)$, which clearly
depends on the traffic rate of the node. It follows that the vector
of node traffic generation probabilities $Q$ is the solution of a
system of flow balance equations $ Q = Q\, {\textbf{T}} +
\lambda$. In steady state, we solve the equation and have
\begin{align}\label{eq:q}
Q = \lambda\, [\textbf{I} - {\textbf{T}}]^{-1}  \,,
\end{align}
where \textbf{I} is the identity matrix and $^{-1}$ denotes the
matrix inversion, which always exists because $\textbf{T}$ is
a probability matrix and thus has eigenvalues strictly less than one \cite{matrix}. Eq.~\eqref{eq:q} gives the fundamental relation
between the packet generation rate, the effect of routing (via
$\textbf{M}$) and the performance of the MAC layer (via the link
reliability $R_{i,j}$). Eq.~\eqref{eq:q} together with the
expressions for the per-link reliability $R_{i,j}$ gives the distribution of the
traffic in the network. In practice, Eq.~\eqref{eq:q} is the
fundamental equation to model mathematically in a simple yet effective manner the joint effects of
MAC and routing. The $i$-th component of the vector $Q$ is the
amount of traffic that the $i$-th node has to forward to its parent per
unit time. This traffic is handled by the MAC and thus
determines the per-link performance such as reliability, delay
and energy consumption.
We remark that due to the acyclic structure of the communication graph,
the elements $Q_i$ can be calculated locally at node $V_i$ by using information on the own traffic
$\lambda_i$, and the estimate of the total coming traffic from children nodes.
When node $V_i$ switches its selected parent to $V_j$, it combines the information on the
traffic $Q_j$, which can be encapsulated in the DIO messages of $V_j$ and the forwarded traffic $V_i$.
The end-to-end reliability of a
node is the product of each link reliability in the path to the
root. Similarly, the end-to-end delay is the sum of the delays in
the path from the transmitter to the root node.
Therefore, we can define routing metrics that exploit our model to locally estimate the reliability and the effects of routing decisions on the traffic distribution, as we show in the next section.

\section{MAC-aware Routing Metrics} \label{sec:mac-aware}
In this section, we present two metrics that are based on the link performance at MAC layer. Moreover, they are simple and easy-to-implement in practice using mechanisms defined in standards.

First, we introduce the $R$-metric. For node $V_i$, we define the metric $R(i) = R_{i,0},$
where  $R_{i,0}$ is the end-to-end reliability between node $V_i$ and $V_0$.
Then, nodes forward their packets by selecting a parent $V_j$ such that the
end-to-end reliability is maximized, i.e.,
$$\underset{j \in \Gamma_i}{\text{maximize}} \qquad R_{i,j} \cdot R(j).$$
 The set of candidate receivers $\Gamma_i$ is composed by the set of nodes that can guarantee a progress towards the destination $V_0$, according to RPL specifications.
 The metric is calculated for all nodes starting from the root $V_0$ and progressing from parents to children in the DODAG.
The $R$-metric extends the ETX metric at the MAC layer, by considering also packet losses due to the MAC contention.
In fact, the ETX is based on the expected number of retransmissions needed to reach a destination, by counting the number of transmissions and ACKs. ETX is an additive metric over the path. $R$-metric is based on the probability that a packet is correctly received in each link of the paths, within a maximum number of backoffs and retransmissions at the MAC layer.

Moreover, the $R$-metric extends the concept of link quality, by including the effects of contention at the MAC layer. In fact, even in case of same LQI indicator among different links, the routing decision determines a different distribution of the traffic over the network and a different level of contention at MAC layer for the forwarding nodes, thus different busy channel probabilities $\alpha_i$, which are included in the expression of the reliability.
We recall that the estimation of the busy channel probability can be performed at the node without extra information needed or modification to the standard IEEE 802.15.4 MAC. Moreover, its estimation is faster than the ETX estimation, which is performed over a certain number of received ACKs, as we report in our experimental evaluation.

For low power applications the reliability can be just set in terms of minimum requirement, and the objective is mainly the network lifetime.
We then propose a metric called $Q$-metric, which distributes the forwarded traffic to provide load balancing in the network.
In particular, the $Q$-metric at nodes $V_i$ computes the traffic $Q_i$. Node $V_i$ selects the forwarding parent by solving the following optimization problem:
\begin{align}\underset{j \in \Gamma_i}{\text{minimize}} \qquad  & P_{t,j} Q_j  + P_{r,j} (Q_j - \lambda_j)  \label{eq:opt}\\
  \text{subject to} \qquad & R_{i,j} \cdot R(j) \geq R_{\min}\,, \nonumber
\end{align}
where $P_{t,j}$ is the power consumption in transmission, and $P_{r,j}$ is the power consumption in reception,  and $R_{\min}$ is the constraint on the reliability required by the application.
The cost function in Eq.~\eqref{eq:opt} is the sum of the cost for transmitting the total  traffic $Q_j$ and cost for receiving traffic generated by children nodes ($Q_j-\lambda_j$).
The values of power consumptions in standard conditions $P_{t,j}$, $P_{r,j}$ can be found in~\cite{Fourty}.
The metric provides load balancing in terms of generated and forwarded traffic.
As far as the implementation of this metric is concerned, node $V_i$ needs only local information about its own forwarded traffic $Q_i$, and the generated and forwarded traffic from each candidate destination, which is available through the exchange of DIO messages.
We recall that load balancing is achieved typically by considering node queues, as in BCP~\cite{BCP}. The back-pressure algorithm in~\cite{BCP}  uses a weighted ETX cost, which includes the queue differential between transmitter and receiver.
The protocol guarantees load balancing by avoiding at any time that node queues overloading. However, the back-pressure metric is not able to capture the contention level when the traffic load is low (which is the case in most WSN applications in real life). In other words, BCP is efficient in saturated cases where the forwarded traffic is high.  On the other hand, the $Q$-metric is able to directly measure the contention level without measuring the node queues and adapt the routing decisions accordingly.
We present the effectiveness of the proposed metrics and a comparison to the back-pressure routing in Section~\ref{exp:res} through experiments.

\section{Experimental Evaluation} ~\label{exp:res}
In this section, we present experimental results related to the
performance of IETF RPL, the contention-based IEEE
802.15.4 MAC and the proposed enhancements.
As a benchmark, we evaluate the performance of our metrics against the back-pressure algorithm proposed in~\cite{BCP}.

We assume that nodes are deployed and connected to form DODAGs as for the topologies in Fig.~\ref{fig:topology}.
We constructed the topology in Fig.~\ref{fig:topology}a to validate the proposed metric in a simple two-hops scenario.
However, we test our metrics also on randomly generated topologies with larger number of nodes as in Fig.~\ref{fig:topology}b, to prove that the results we present in
this section give general insights and can be derived for different multi-hop topologies.

The IEEE 802.15.4 protocol is implemented on a test-bed
using the TelosB platform~\cite{telosb} running the Contiki OS~\cite{contiki}. We assume that each node generates the same
traffic with rate $\lambda=[0.1 \div 10]$ pkt/s, except $V_2$ that generates traffic with rate $\lambda_2=20$ pkt/s (dominant node). We chose the unslotted MAC modality since it is one of the
recommended in the IETF RPL standard. However, the methodology that we
have proposed above can
be applied to any randomized MAC, compatible with IETF RPL.
We represent then a realistic network operation in which heterogeneous traffic conditions are determined in the network both by a different traffic generation rate among nodes and distribution of the forwarded traffic among various routing paths.

\begin{figure}[t!] \centering
\includegraphics[width=0.7\textwidth]{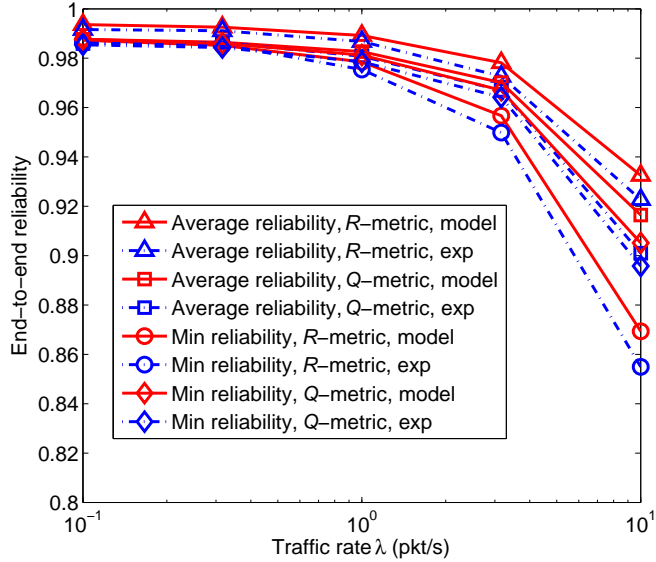}
\caption{End-to-end reliability for the
multi-hop topology in Figure~\ref{fig:topology}. Average reliability is evaluated among nodes $V_4$ to $V_7$.
Minimum reliability is the average reliability of the worst case path. \label{fig:mh_rel_mag}}
\end{figure}
\begin{figure}[t] \centering
\includegraphics[width=0.7\textwidth]{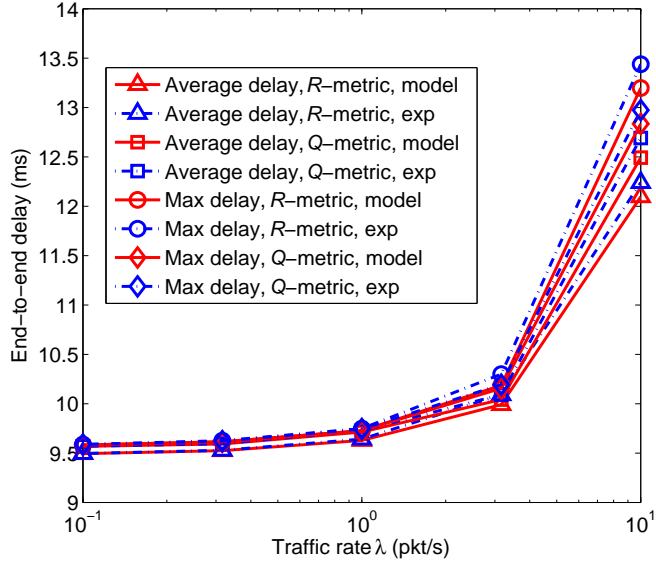}
\caption{End-to-end delay for the multi-hop
topology in Figure~\ref{fig:topology}. Average delay is evaluated among nodes $V_4$ to $V_7$.
Maximum delay is the average delay of the worst case path.\label{fig:mh_lat_mag}}
\end{figure}


\subsection{Model Validation}

In the first set of experiments, we validate our analytical model against experimental results of the two proposed metrics for the topology in Fig.~\ref{fig:topology}a.
Fig.~\ref{fig:mh_rel_mag} shows the end-to-end reliability vs the traffic rate as obtained
by the mathematical model and the experiments when our proposed $R$-metric and $Q$-metric are used.
We show the average reliability among nodes $V_4$ to $V_7$ and
the minimum path reliability achieved in the network. We observe that the experimental results are very close to the analytic results.
In Fig.~\ref{fig:mh_rel_mag}, the minimum reliability for the $R$-metric is achieved for the path that includes $V_4$ and $V_1$.
The reliability with the $Q$-metric does not vary significantly in the paths and the minimum reliability is only slightly lower than the average reliability and, furthermore, it is greater than the minimum reliability for the $R$-metric. In our experimental evaluation, the gap in the minimum reliability is around 5$\%$ for $\lambda=10$ pkt/s and it increases as the traffic rate increases. We notice that, even though the $R$-metric achieves the best average performance from a network perspective, the $Q$-metric is preferable if a guaranteed reliability is required for all paths in the network, which is often desired by IoT applications \cite{palattella_iot}.

In Fig.~\ref{fig:mh_lat_mag}, the end-to-end delay as obtained
by the mathematical model and the experiments with $R$-metric and $Q$-metrics is presented.
We visualize the average delay achieved for nodes $V_4$--$V_7$ and
the maximum path delay in the network.
By reducing the level of contention at the MAC layer, the average delay is lower when the $R$-metric is used.
In the range of traffic rates analyzed, which is of interest for RPL applications~\cite{ROLL}, the queueing delay does not influence the performance significantly. Therefore the $R$-metric guarantees also a minimization of the average delay. However, the maximum delay, which is again achieved  for the path that includes $V_4$ and $V_1$,  is lower with the $Q$-metric.  If there are delay deadlines for all nodes, as in the proposed network scenario, the $Q$-metric is preferable. The gain with the $Q$-metric is 4$\%$ when compared to the $R$-metric for  $\lambda=10$ pkt/s.

\subsection{Performance Comparison - ETX}

In this section, we show how the $R$-metric outperforms ETX by accounting for MAC-routing interactions.
Consider the multi-hop topology in Fig.~\ref{fig:topology}a. Node $V_7$ has two paths to the destination, one path through $V_2$ and the other through $V_3$. Assume that the path through $V_2$ has ETX$_{7,2}=2.1$ and ETX$_{2,0}=2.1$, which determines a total expected number of retransmissions ETX$_{7,2,0}=4.2$ to the destination. The second path has ETX$_{7,3}=1.1$ and ETX$_{3,0}=2.9$, which makes a total ETX$_{7,3,0}=4.0$.
In absence of a retry limit at MAC layer, the second path through $V_3$ has the minimum ETX value and gives the highest end-to-end delivery ratio. However, if we set a maximum number of retransmissions $n=4$, as specified by the IEEE 802.15.4 standard, and by assuming independent loss probability for consecutive retransmissions, the end-to-end success rate is $92.3\%$ in the path through $V_2$, while only $87.7\%$ in the path through $V_3$, due to higher packet loss probability in the link $(3,0)$. The path through $V_2$ has a $5\%$ worse ETX value but it guarantees a $5\%$ better end-to-end reliability, and this effect is expected to be more evident by considering correlation between consecutive retransmissions.

\begin{figure}\centering
  \includegraphics[width=0.7\linewidth]{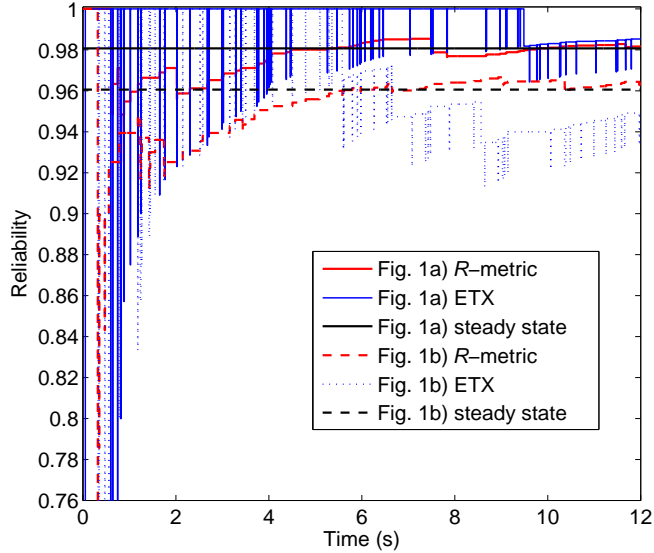}
  \caption{Reliability estimation with R-metric and ETX metric for topologies in Fig.~\ref{fig:topology}a and in Fig.~\ref{fig:topology}b.}\label{fig:estimation}
\end{figure}

In Fig.~\ref{fig:estimation}, we report the estimated link reliability by using $R$-metric  and ETX metric for the two topologies in Fig.~\ref{fig:topology}.
The steady state value represent the average reliability in stationary conditions after sending $10^5$ packets.
The ACK-based estimation used in ETX shows larger variability and slower convergence speed with respect to the $R$-metric mechanism bases on the busy channel probability.

\subsection{Performance Comparison - Backpressure Routing}

\begin{figure}[h] \centering
\includegraphics[width=0.7\textwidth]{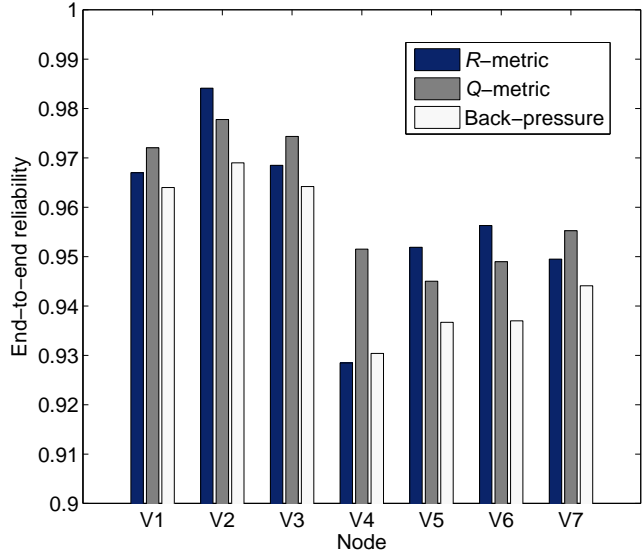}
\caption{End-to-end node reliability for the multi-hop
topology in Fig.~\ref{fig:topology}a, by fixing $\lambda_i=5$ pkt/s for $i\neq2$ and $\lambda_2=20$ pkt/s.\label{fig:mh_rel_node}}
\end{figure}

 \begin{figure}[h] \centering
\includegraphics[width=0.7 \linewidth]{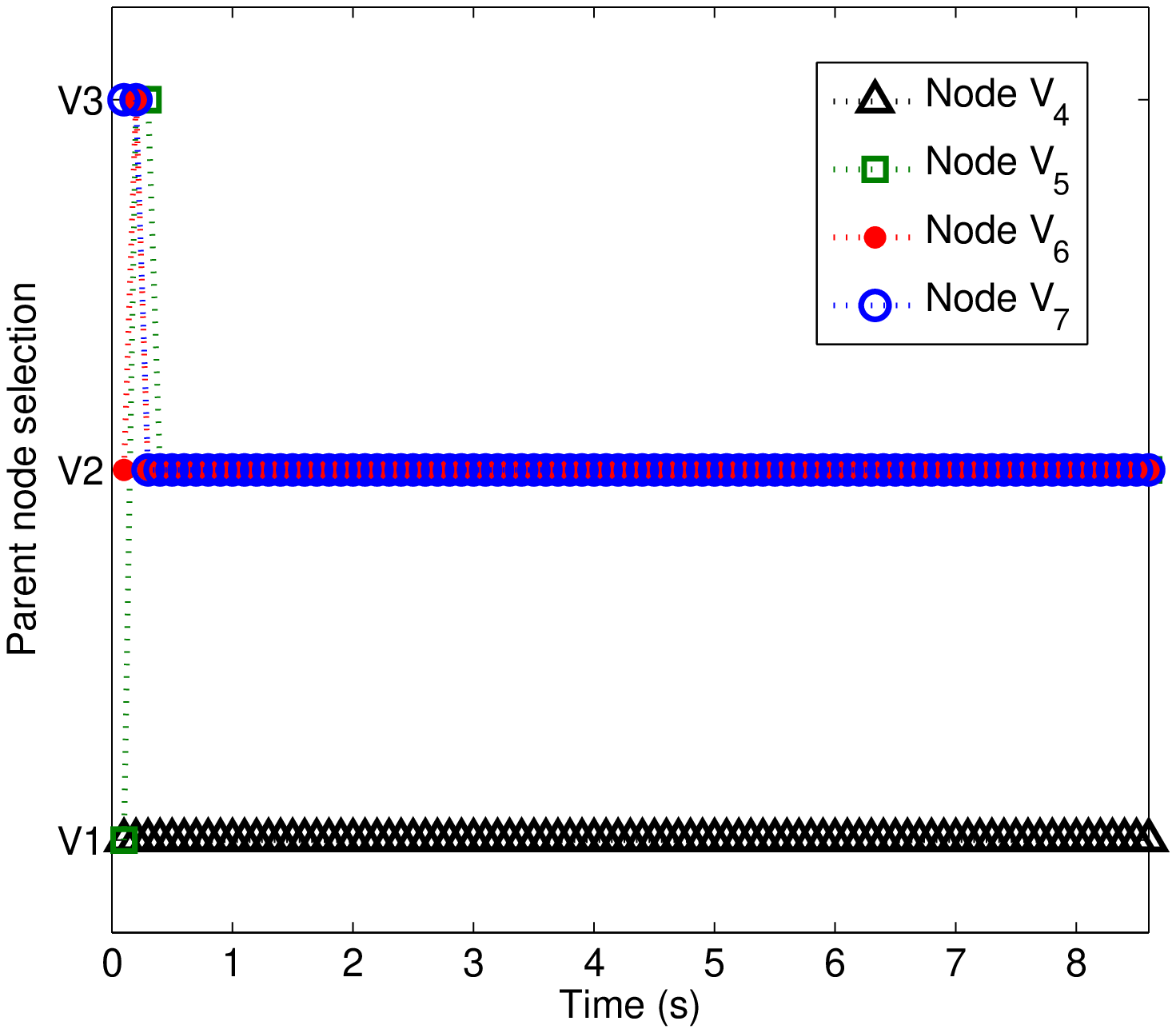}
\caption{Parent node selection vs. time for $R$-metric in the multi-hop topology in Fig.~\ref{fig:topology}a by fixing $\lambda_i=10$ pkt/s for $i\neq2$ and $\lambda_2=20$ pkt/s.
 \label{fig:route_r}}
\end{figure}

 \begin{figure}[h] \centering
\includegraphics[width=0.7 \linewidth]{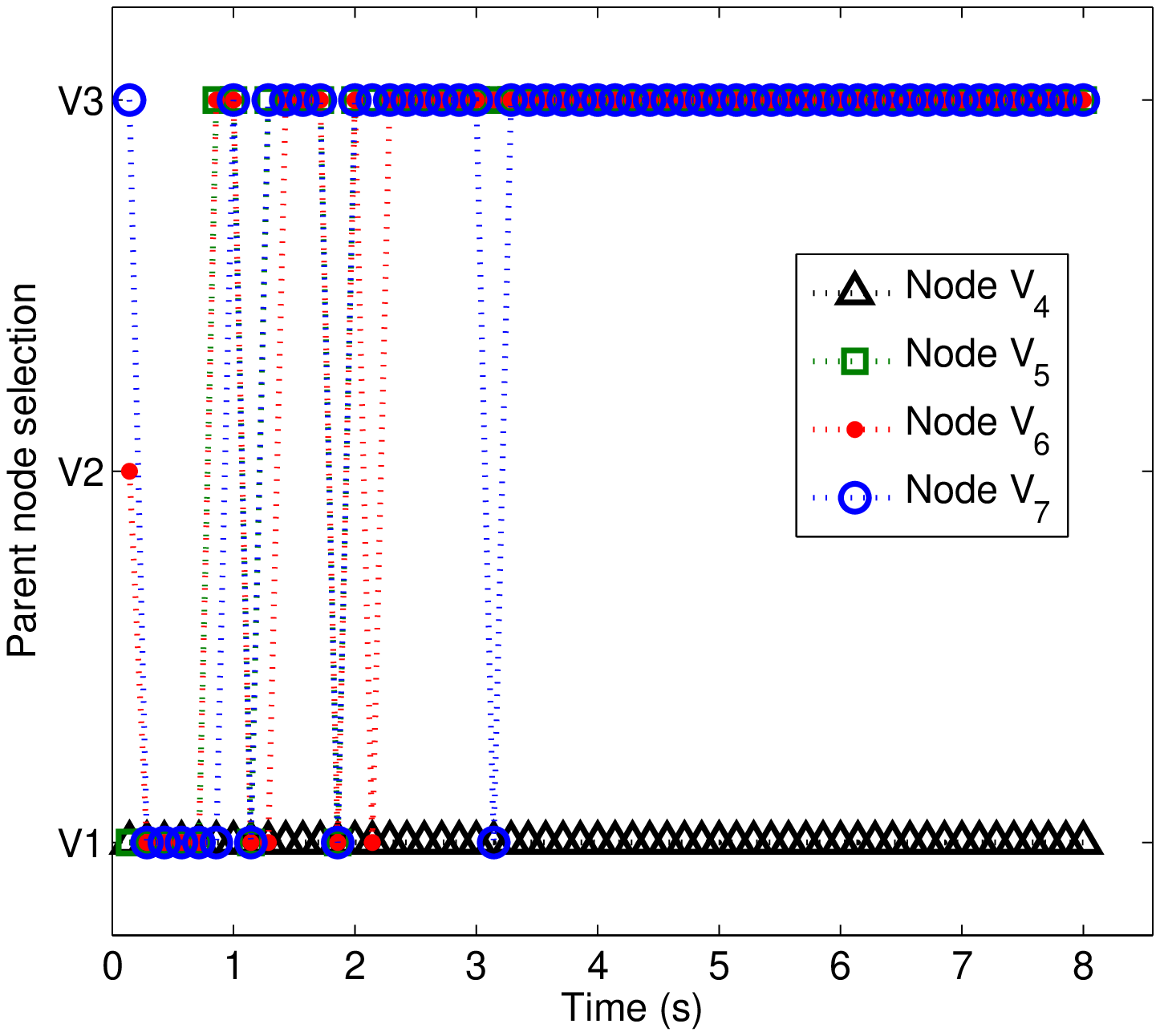}
\caption{Parent node selection vs. time for $Q$-metric in the multi-hop topology in Fig.~\ref{fig:topology}a by fixing $\lambda_i=10$ pkt/s for $i\neq2$ and $\lambda_2=20$ pkt/s.
 \label{fig:route_q}}
\end{figure}

 \begin{figure}[h] \centering
\includegraphics[width=0.7 \linewidth]{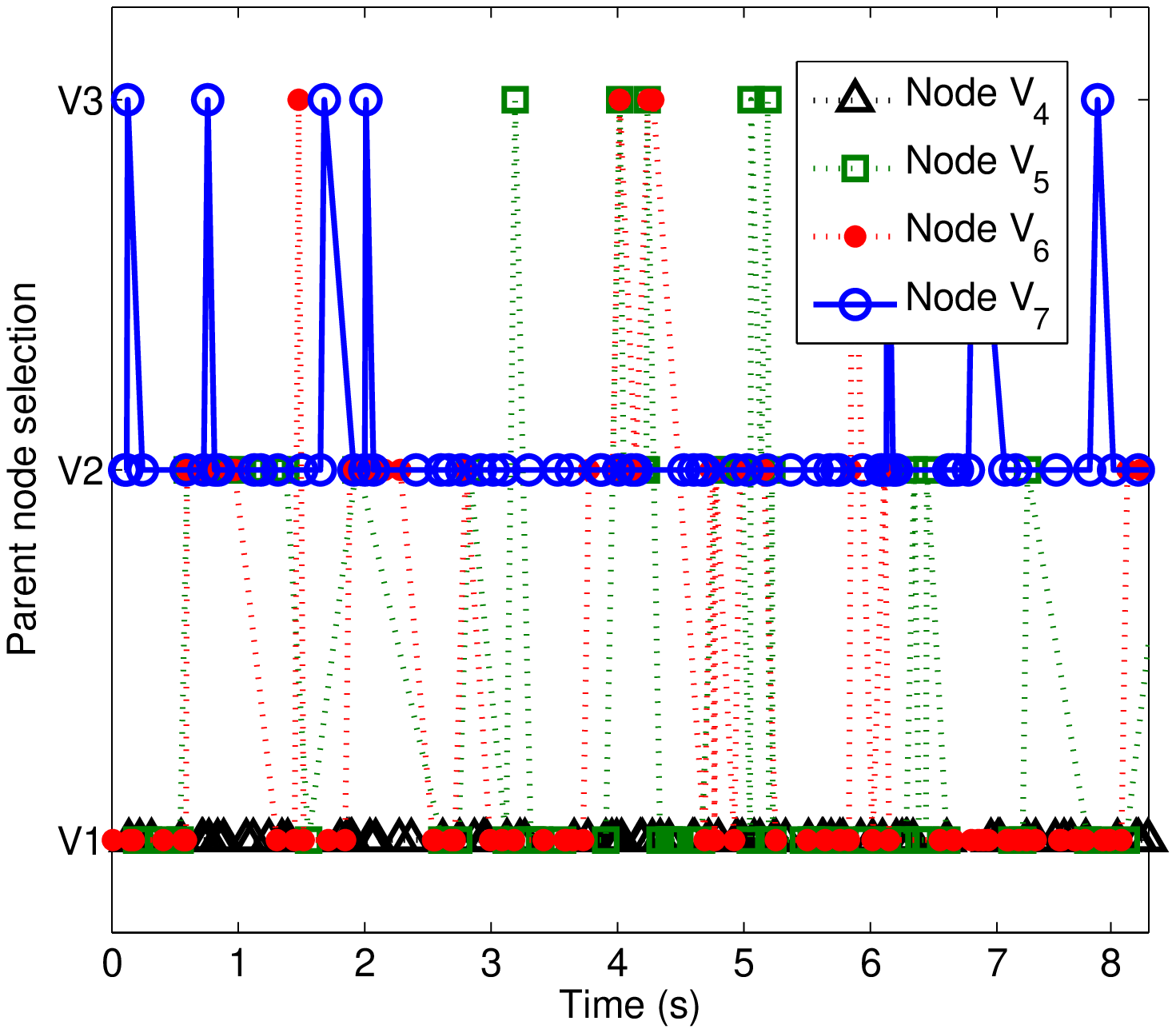}
\caption{Parent node selection vs. time for back-pressure routing in the multi-hop topology in Fig.~\ref{fig:topology}a by fixing $\lambda_i=10$ pkt/s for $i\neq2$ and $\lambda_2=20$ pkt/s.
 \label{fig:route_back}}
\end{figure}

In Fig.~\ref{fig:mh_rel_node}, we show the end-to-end reliability of each node of Fig.~\ref{fig:topology}a, by fixing $\lambda_i=5$ pkt/s for $i\neq2$ and $\lambda_2=20$ pkt/s. We compare $R$-metric, $Q$-metric, and back-pressure. The $R$-metric guarantees high reliability for the dominant node $V_2$ which forwards most of the traffic in the network. However, as anticipated from the results in Fig.~\ref{fig:mh_rel_mag}, the reliability of $V_4$ is compromised. The $Q$-metric guarantees the reliability constraint in $V_4$ and outperforms the back-pressure metric. A frequent parent switching in the back-pressure routing determines an increase of the traffic due to high DIO message transmissions that affect the reliability.

To better  understand the results shown in Fig.~\ref{fig:mh_rel_node}, it is necessary to study how the two routing metrics distribute the traffic among the various paths.
In Figs.~\ref{fig:route_r}~--~\ref{fig:route_back}, we plot the time evolution of the parent selection for each end-device in the network by using $R$-metric, $Q$-metric, and back-pressure respectively. In the experiment, we set $\lambda_i=10$ pkt/s for $i\neq2$ and $\lambda_2=20$ pkt/s. The end-devices start from a random initial condition and explore the various routing paths to determine the next-hop node according to the selected metric. When using the $R$-metric, nodes $V_4$--$V_7$ tend to forward their traffic through the dominant node
$V_2$, which generates traffic $\lambda_2=20$ pkt/s, thus reducing the level of contention at the MAC layer.
The level of contention is measured as the probability that the channel is occupied by concurrent transmissions at the same time.
Therefore, by directing the forwarded traffic to a single node, the level of contention reduces.
When using the $Q$-metric,  nodes $V_4$ to $V_7$ tend to distribute the traffic uniformly in the set of candidate receivers $V_1$ to $V_3$ thus increasing the level of contention at MAC layer. The average end-to-end reliability of the network is then higher for the $R$-metric. However, by reducing the level of contention at the dominant nodes, the $R$-metric increases the level of contention for the communication paths that do not include the dominant nodes. Therefore, the end-to-end reliability in the path that includes $V_4$ and $V_1$ is affected significantly by the dominant node $V_2$.
When using back-pressure routing, we notice frequent switches in the parent node selection. Nodes choose their parent on a packet base, by looking at the current queues. However, due to the unsaturated traffic, the value of the queues oscillates between 0 and 1 among the nodes at each transmission, with sporadic cases in which the queues are affected by the unbalanced traffic.

%

\begin{figure}[h] \centering
\includegraphics[width=0.7\textwidth]{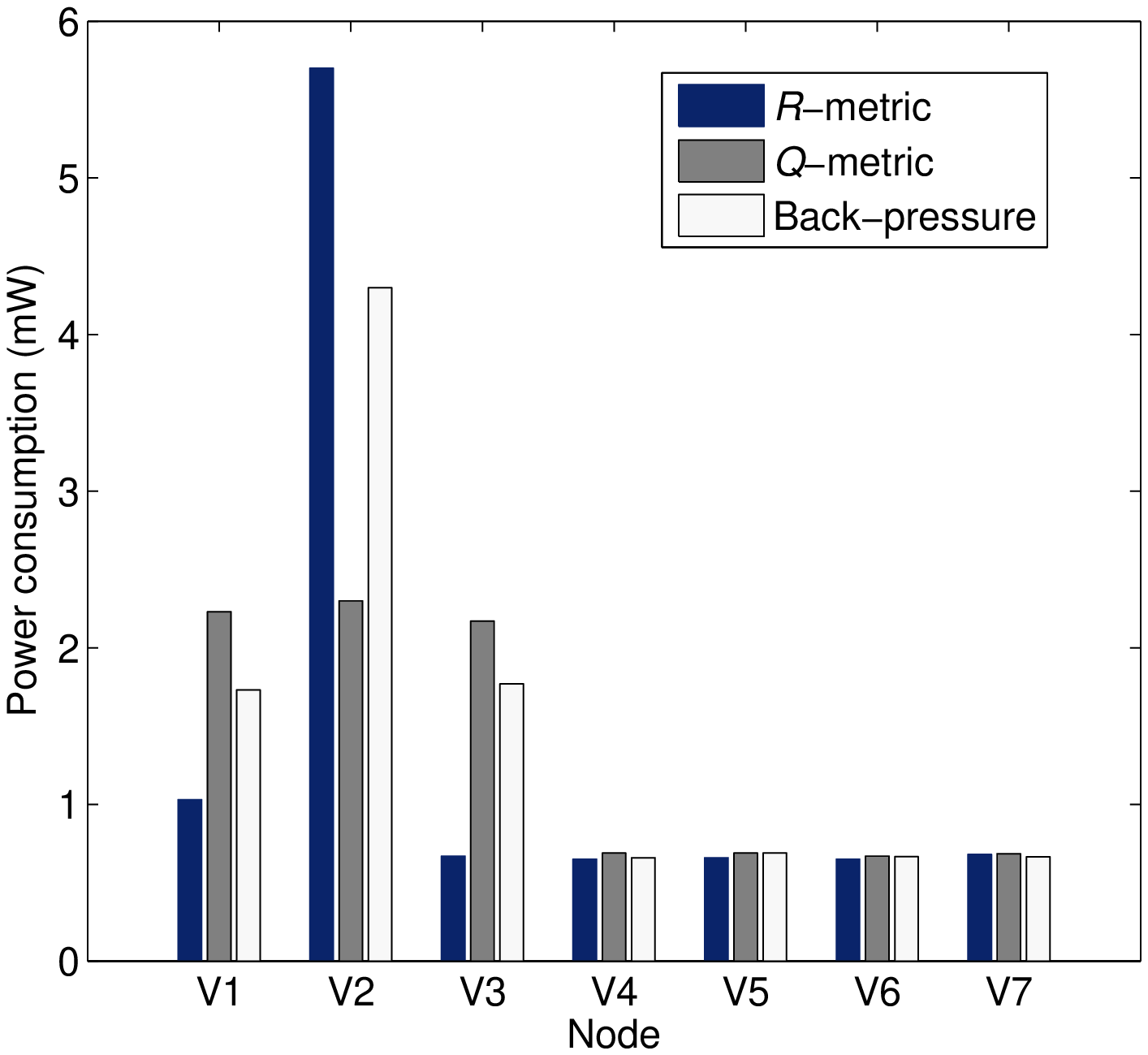}
\caption{Average node power consumption for the multi-hop
topology in Fig.~\ref{fig:topology}a, by fixing $\lambda_i=5$ pkt/s for $i\neq2$ and $\lambda_2=20$ pkt/s.\label{fig:mh_ene_node}}
\end{figure}

\begin{figure}[h!] \centering
\includegraphics[width=0.7\textwidth]{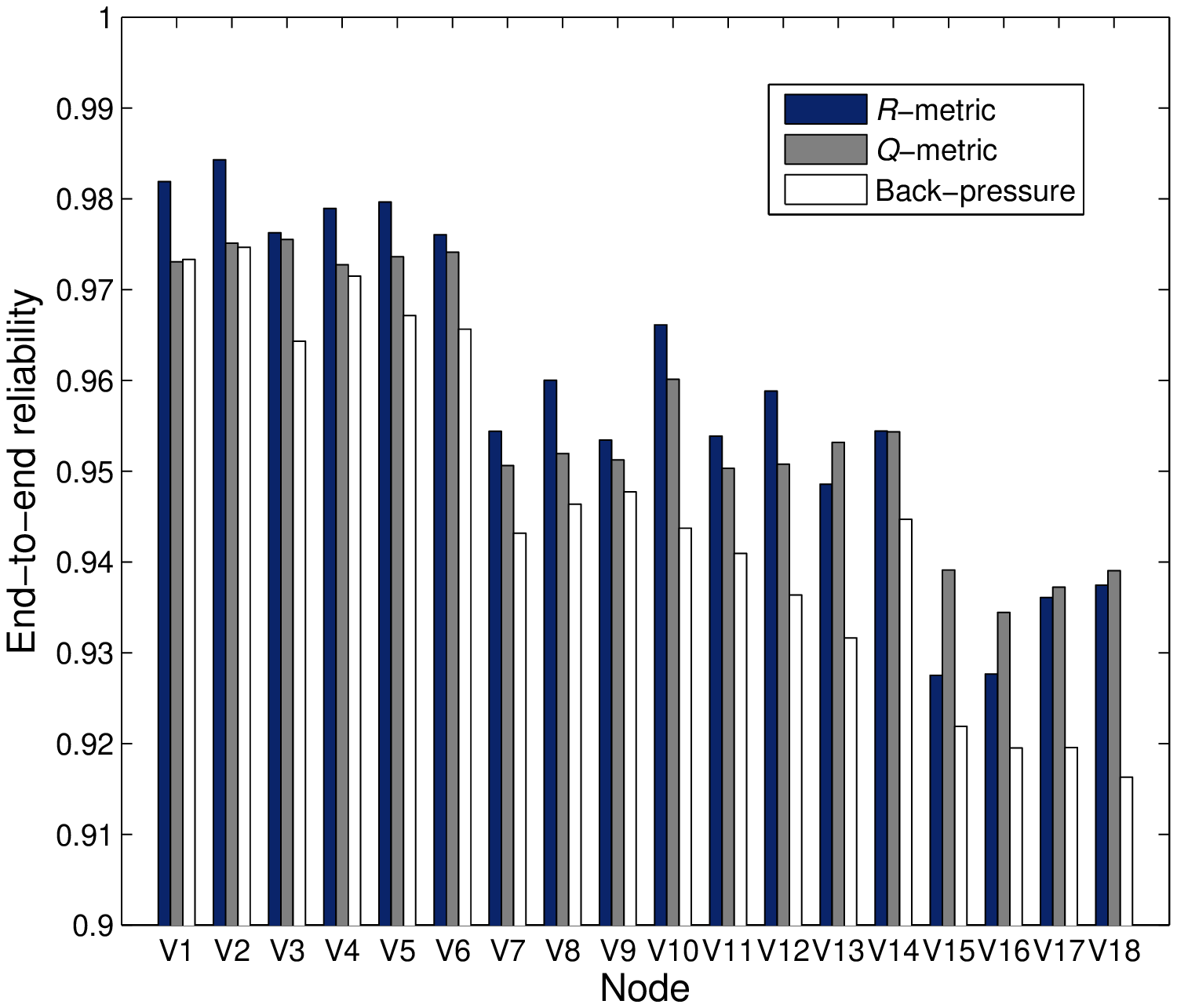}
\caption{End-to-end node reliability for the multi-hop
topology in Fig.~\ref{fig:topology}b, by fixing $\lambda_i=1$ pkt/s for $i=1,...,N$.\label{fig:mh_rel_node20}}
\includegraphics[width=0.7\textwidth]{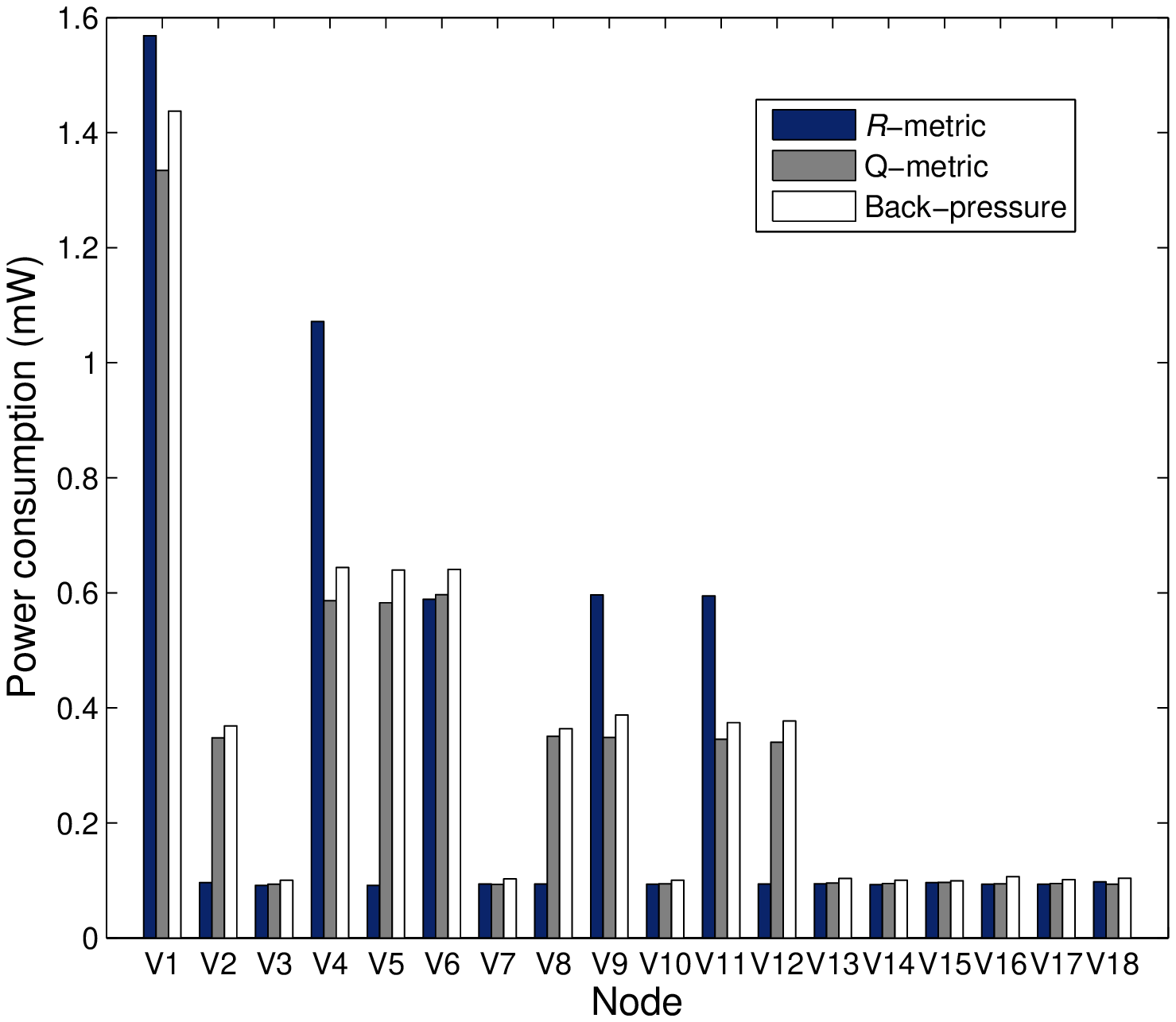}
\caption{End-to-end node reliability for the multi-hop
topology in Fig.~\ref{fig:topology}b, by fixing $\lambda_i=1$ pkt/s for $i=1,...,N$.\label{fig:mh_pow_node20}}
\end{figure}

In Fig.~\ref{fig:mh_ene_node}, we show the average power consumption of each node, by fixing $\lambda_i=5$ pkt/s for $i\neq2$ and $\lambda_2=20$ pkt/s.
The power consumption is calculated by considering the sum of the contributions in transmission, reception, idle-listening, and carrier sensing for each node.
By choosing the dominant node $V_2$ as forwarder, the $R$-metric determines an unbalanced energy consumption. Node $V_2$ has a power consumption up to $6$ mW, while the rest of the network operates between $0.5$ mW and $1$ mW. With the $Q$-metric, the power consumption is more balanced among nodes and the maximum consumption, which is crucial for the network lifetime, decreases of at least a factor 2 respect to the $R$-metric. The back-pressure routing present a reduction of the maximum energy consumption with respect to the $R$-metric. However, the dominant node $V_2$ consumes $70\%$ more power compared to the $Q$-metric.

The simulations above were replicated also for the topology in Fig.~\ref{fig:topology}b.
In Fig.~\ref{fig:mh_rel_node20}, we show the end-to-end reliability of each node, by fixing $\lambda_i=1$ pkt/s for $i=1,...,N$. We compare $R$-metric, $Q$-metric, and back-pressure. Both $R$-metric and $Q$-metric outperform the back-pressure metric in terms of end-to-end reliability. A lower variance among nodes is revealed for the $Q$-metric, while the $R$-metric maximizes the reliability of dominant paths.

In Fig.~\ref{fig:mh_pow_node20}, we show the average power consumption of each node, by fixing $\lambda_i=1$ pkt/s for $i=1,...,N$. We compare $R$-metric, $Q$-metric, and back-pressure.
The $Q$-metric guarantees more balanced load distribution and the maximum power consumption, which is observed in $V_1$, is reduced of 15$\%$ with respect to $R$-metric, and 7$\%$ with respect to back-pressure.

\begin{figure}[h!]\centering
  \includegraphics[width=0.7\textwidth]{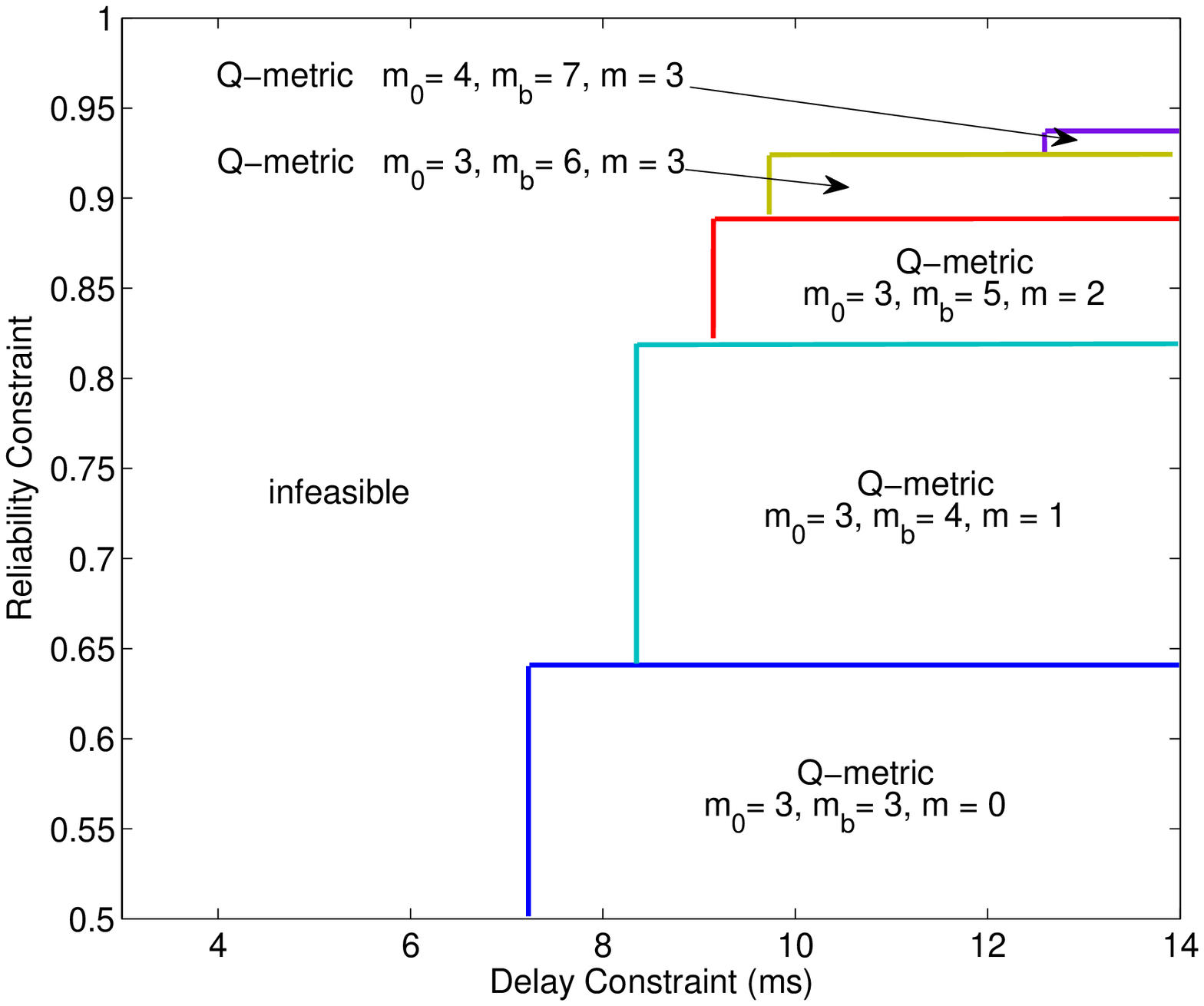}
  \caption{IEEE 802.15.4 MAC and RPL protocol parameters selection for the multi-hop topology in Fig.~\ref{fig:topology}a by fixing $\lambda_i=5$ pkt/s for $i\neq2$ and $\lambda_2=20$ pkt/s.}\label{fig:selection}
\end{figure}

\subsection{Protocol Parameters Selection}

Now we turn our attention to show how complicated and inefficient the selection of the MAC parameters or routing metrics can be, if the mutual interactions are not considered.
We report the results of a mathematical tool for
parameter selection, by considering the multi-hop topology in Fig.~\ref{fig:topology}a.
In the tool, we include the analytical model of the IEEE 802.15.4 MAC, derived in~\cite{PG_TVT},
and the analysis of the interaction with IETF RPL developed in this paper.
The output of the tool is defined as the protocol and the set of MAC-routing parameters
that maximize the network lifetime for certain reliability
and delay constraints imposed by the application to all nodes, as reported
on the x and y axis of Fig.~\ref{fig:selection}.

We consider the unslotted IEEE
802.15.4 MAC and we let the protocol selection mechanism choose the
initial backoff exponent $m_{0}=[3\div8]$, the maximum backoff exponent
$m_{b}=[m_0\div8]$, and the maximum number of backoffs $m = [0\div4]$.
Moreover, we apply RPL and we let the mechanism choose between $R$-metric and $Q$-metric.
The traffic rate is $\lambda_i=5$ pkt/s for $i\neq2$ and $\lambda_2=20$ pkt/s.
The $Q$-metric is always preferred to the $R$-metric, whenever the solution is feasible.
This is compliant with the analysis and experiments presented in the previous sections,
since the constraints are for all nodes, and the objective is the minimization of the energy consumption
of the dominant node.
In general, an increase of the MAC parameters determines an increase in the energy consumption and in the delay.
However, the reliability increases too.
For a reliability constraint smaller than $65\%$, and delay constraint greater than $7.5$ ms,
the optimal MAC parameters are $m_{0}=3$, $m_{b}=3$, and
$m=0$.  However, the optimal parameters increase as the reliability constraint become stricter,  as the solutions become unfeasible.
For a reliability constraint above $90\%$,  the optimal solution is obtained by increasing both the number of backoffs $m$ and the backoff windows $m_{0}$ and $m_{b}$.
The node energy consumption associated to $m_{0}=3$, $m_{b}=3$, and
$m=0$ is about $20\%$ lower than the consumption with default parameters $m_{0}=3$, $m_{b}=8$, and $m=4$. In addition, as we showed in Fig.~\ref{fig:mh_ene_node}, the maximum energy consumption is halved as we choose the $Q$-metric over the $R$-metric.
Therefore, by optimally selecting routing metric and MAC parameters according to the reliability and delay constraints, it is possible to obtain a significant impact on the performance.


\section{Conclusions} \label{sec:conclusions}
In this paper, an analysis to characterize the complex inter-dependence among the basic MAC and routing protocols in IoT was presented. Moreover, a mathematical framework for joint
optimization of the MAC and the routing layers parameters was proposed to enhance the existing standards.
Specifically, novel metrics that take into account the
dynamic behavior of the MAC and routing layers were introduced: $R$-metric and $Q$-metric. An extensive comparison of the existing ETX metric with the $R$-metric, which considers both the level of contention and the protocol parameters, was performed.
It was shown that the $R$-metric achieves high average link reliability. However, it is not able to provide balanced reliability in the network.
The $Q$-metric was proposed, having in mind that a minimum reliability or a maximum delay is required for all the nodes. Extended experiments, where the proposed metrics were compared to the existing back-pressure routing, supported our mathematical analysis.
The inclusion of such an analysis and an experimental study in the current
standardization process could be very beneficial in the direction of improving the
performance of IoT protocols under realistic conditions.

\section*{Acknowledgements}
The work of the authors was supported by the EU projects Hycon2, Hydrobionets and the Swedish Research Council.

\section*{References}
\bibliographystyle{IEEEtran}
\bibliography{ref}

\end{document}